\documentclass[prl,amsmath,amssymb,twocolumn]{revtex4-2}
\usepackage{graphicx}
\usepackage{adjustbox}
\usepackage{bm}
\usepackage{color}
\usepackage{braket}
\usepackage{standalone}
\usepackage{multirow}
\usepackage{tikz}
\usepackage{mathrsfs}
\usepackage{dsfont}
\usepackage{physics}
\usepackage[colorlinks,bookmarks=true,citecolor=blue,linkcolor=blue,urlcolor=blue]{hyperref}
\usepackage{cleveref}

\begin{document}

\title{Braid Protected Topological Band Structures with Unpaired Exceptional Points}

\author{J. Lukas K. K\"onig$^1$, Kang Yang$^{1,2}$, Jan Carl Budich$^3$, and Emil J. Bergholtz$^1$}

\affiliation{$^1$Department of Physics, Stockholm University, AlbaNova University Center, 10691 Stockholm, Sweden\\
$^2$ Dahlem Center for Complex Quantum Systems and Institut f\"ur Theoretische Physik,
Freie Universit\"at Berlin, Arnimallee 14, 14195 Berlin, Germany\\
$^3$Institute of Theoretical Physics, Technische Universit\"at Dresden and W\"urzburg-Dresden Cluster of Excellence ct.qmat, 01062 Dresden, Germany
}
\date{\today}

\begin{abstract}
We demonstrate the existence of topologically stable unpaired exceptional points (EPs), and construct simple non-Hermitian (NH) tight-binding models exemplifying such remarkable nodal phases. While fermion doubling, i.e. the necessity of compensating the topological charge of a stable nodal point by an anti-dote, rules out a direct counterpart of our findings in the realm of Hermitian semimetals, here we derive how noncommuting braids of complex energy levels may stabilize unpaired EPs. Drawing on this insight, we reveal the occurrence of a single, unpaired EP, manifested as a non-Abelian monopole in the Brillouin zone of a minimal three-band model. This third-order degeneracy represents a sweet spot within a larger topological phase that cannot be fully gapped by any local perturbation. Instead, it may only split into simpler (second-order) degeneracies that can only gap out by pairwise annihilation after having moved around inequivalent large circles of the Brillouin zone. Our results imply the incompleteness of a topological classification based on winding numbers, due to non-Abelian representations of the braid group intertwining three or more complex energy levels, and provide insights into the topological robustness of non-Hermitian systems and their non-Abelian phase transitions.

\end{abstract}
\maketitle

Following the experimental discovery of Dirac and Weyl semimetals in solid state materials \cite{novoselov2004electric,PhysRevLett.113.027603,xu2015discoverya,xu2015discoveryb,PhysRevX.5.011029}, Bloch bands with topologically stable nodal points have become a major focus of research far beyond the field of condensed matter physics \cite{RevModPhys.90.015001,novoselov2004electric,PhysRevLett.113.027603,lu2013weyl,lu2014topological,lu2015experimental,PhysRevX.5.011029,PhysRevX.5.031013,xu2015discoverya,xu2015discoveryb,yang2015weyl,huang2016spectroscopic}. 
In crystalline systems, Bloch's theorem requires the periodicity of the band structure in reciprocal space. As a consequence, nodal points carrying a topological charge must be compensated for to allow the eigenstates to seamlessly fit together at the zone boundaries of the first Brillouin zone (BZ). 
A prominent example along these lines is provided by stable Weyl nodes \cite{volovik2003universe,annurev-conmatphys-031016-025458,RevModPhys.90.015001}, which are required to occur in pairs with opposite chirality. 
Under the name fermion doubling, such constraints have been discussed for decades \cite{NIELSEN198120,NIELSEN1981173}.

In dissipative systems described by effective non-Hermitian (NH) operators, exceptional points (EPs) \cite{Berry2004,Heiss_2012,PhysRevX.6.021007,ep-optics,RevModPhys.93.015005,ding2022non} represent the generic counterpart of diagonalizable degeneracies. Since stable EPs may carry topological charge in the sense of a relative winding in the complex energy plane, analogous constraints to the aforementioned fermion doubling have been reported based on $\mathbb Z$ discriminants \cite{PhysRevLett.126.086401}. 
However, as an additional twist, EPs are known to be of intrinsic non-Abelian nature, as characterized by the braid group \cite{PhysRevA.98.023818,PhysRevB.103.155129,PhysRevB.101.205417,PhysRevLett.126.010401,wang2021topological,patil2022measuring}. 
This entails unique NH topological semimetal phases that elude elementary winding number descriptions \cite{PhysRevX.8.031079,PhysRevLett.121.086803,PhysRevX.9.041015,PhysRevLett.127.186602,kawabataClassificationExceptionalPoints2019a}. 
The fusion of nodal points, a phase transition, is topologically path-dependent \cite{RevModPhys.51.591}.

\begin{figure}[h!]
    \centering
    \includegraphics[width=\linewidth]{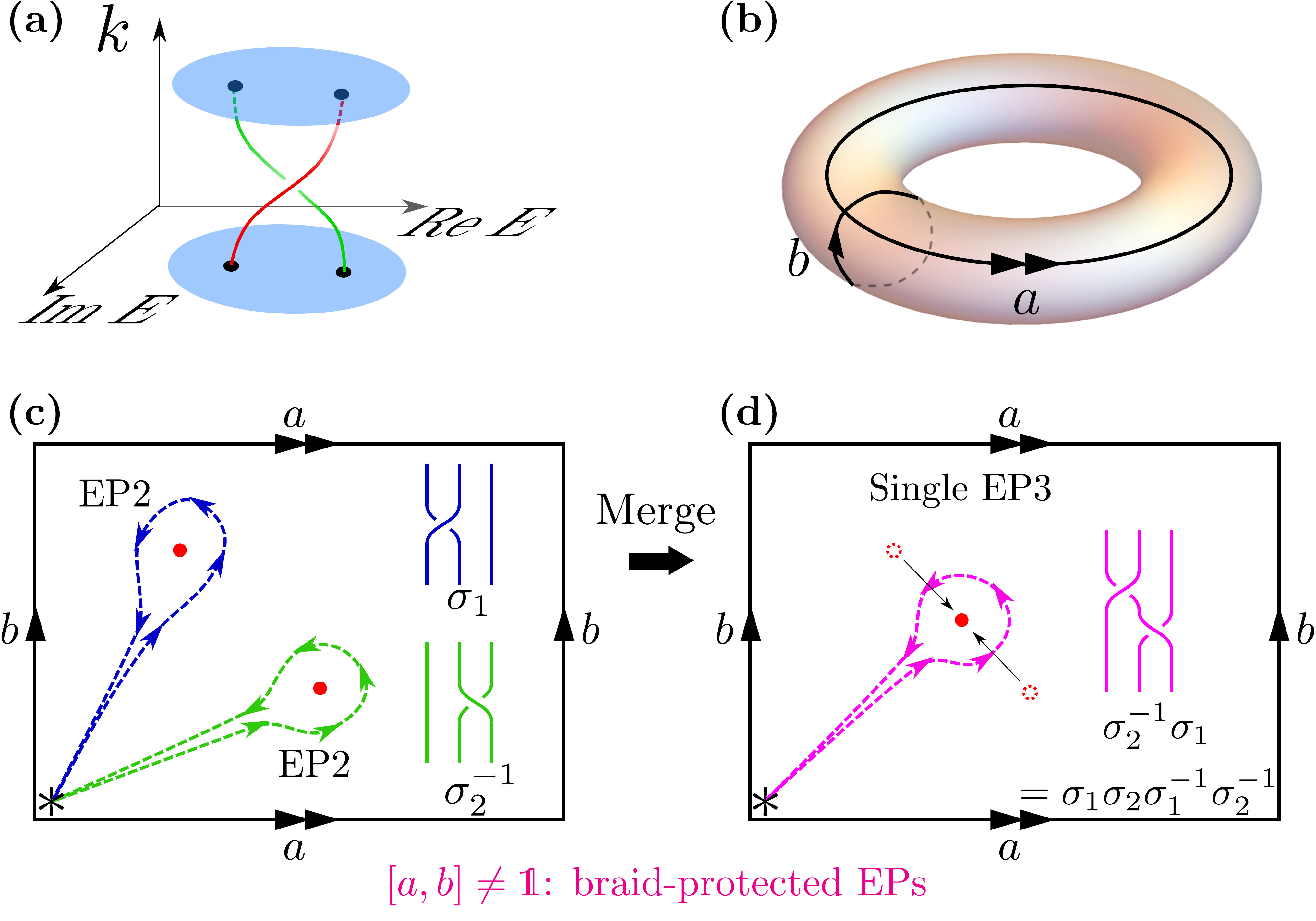}
    \caption{Illustration of key results. 
    \textbf{(a)} The braiding of complex energy levels as a function of lattice momentum is at the heart of our analysis.
    \textbf{(b)} In the two-dimensional Brillouin zone, NH phases are characterized by two braid elements $(a,b)$ along the noncontractible loops. 
    For gapped systems, $a$ and $b$ commute (\emph{group} commutator $aba^{-1}b^{-1}=[a,b]=\mathds{1}$). 
    \textbf{(c)} 
    Minimal model of a band structure with a nontrivial such commutator, hosting two unpaired EP2s.
    \textbf{(d)} While the unpaired EP2s may merge into a single EP3, the nontrivial braid topology implies that the system remains gapless. EPs are denoted by red dots and base points of loops by $\ast$ in all panels.
    }
    \label{fig_sum}
\end{figure}

\begin{figure*}[ht]
	\centering
    \includegraphics[width=\textwidth]{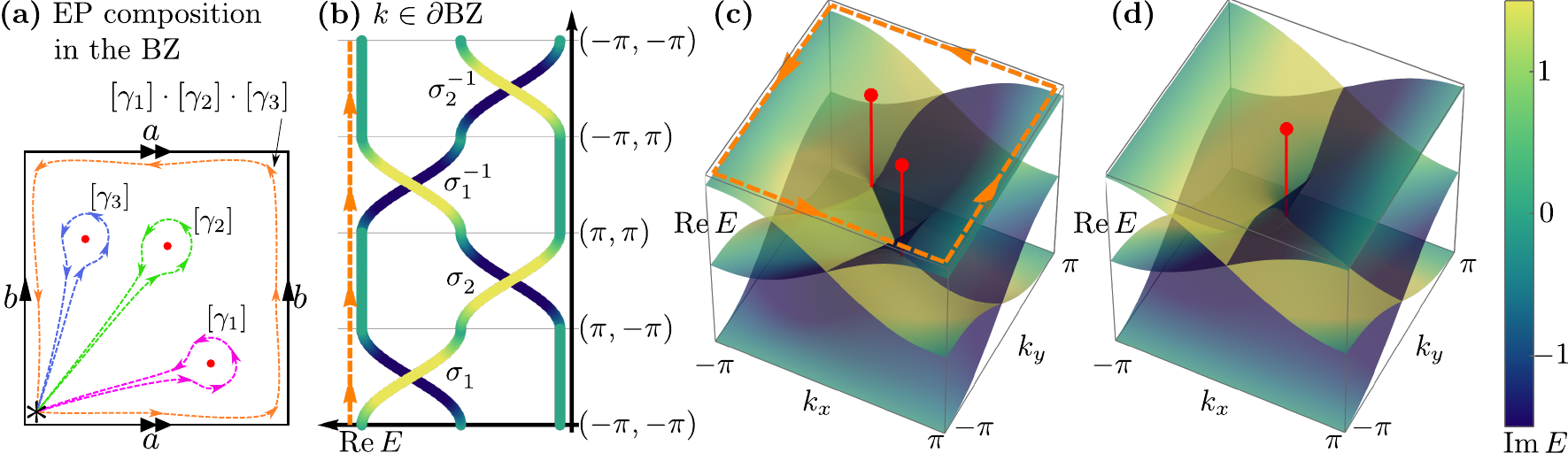}
    \caption{
    Composition rule for braids. \textbf{(a)} Schematic: Braids around degeneracies must add up to the boundary braid.
    \textbf{(b)} Boundary braid:
    Spectrum of Eq.~\eqref{eq:simple-boundary-model} shown along the BZ boundary, counterclockwise from \((-\pi,-\pi)\).
    Associating \(\sigma_i\) to a counterclockwise exchange of eigenvalues \(E_i,E_{i+1}\) in the complex plane allows us to read off the boundary braid as \(\sigma_1^{\vphantom{}} \sigma_2^{\vphantom{}} \sigma_1^{-1} \sigma_2^{-1}\).
    \textbf{(c)}
    Spectrum of Eq.~\eqref{eq:simple-boundary-model} shown on the entire BZ.
    The orange path corresponds to the path traversed in (a) and (b) that gives rise to the boundary braid. 
    The two degeneracies marked in red carry charges \(\sigma_2^{-1}\) and \(\sigma_1\), which add up to the boundary charge [see Eq.~\eqref{eq:braid-word-short}].
    \textbf{(d)}
    Tuning this model to fuse the two EPs [see Eq.~\eqref{eq:tune-to-single}]
    cannot annihilate them if the perturbation conserves the boundary charge, which is why we call these degeneracies unpaired.
    Instead a single nontrivial EP3 remains as a non-Abelian monopole that compensates the entire braid charge of the boundary. 
    }
    \label{fig:ham-d-3D-spectra}
\end{figure*}

\begin{figure*}[!ht]
	\centering
	\includegraphics[width=\linewidth]{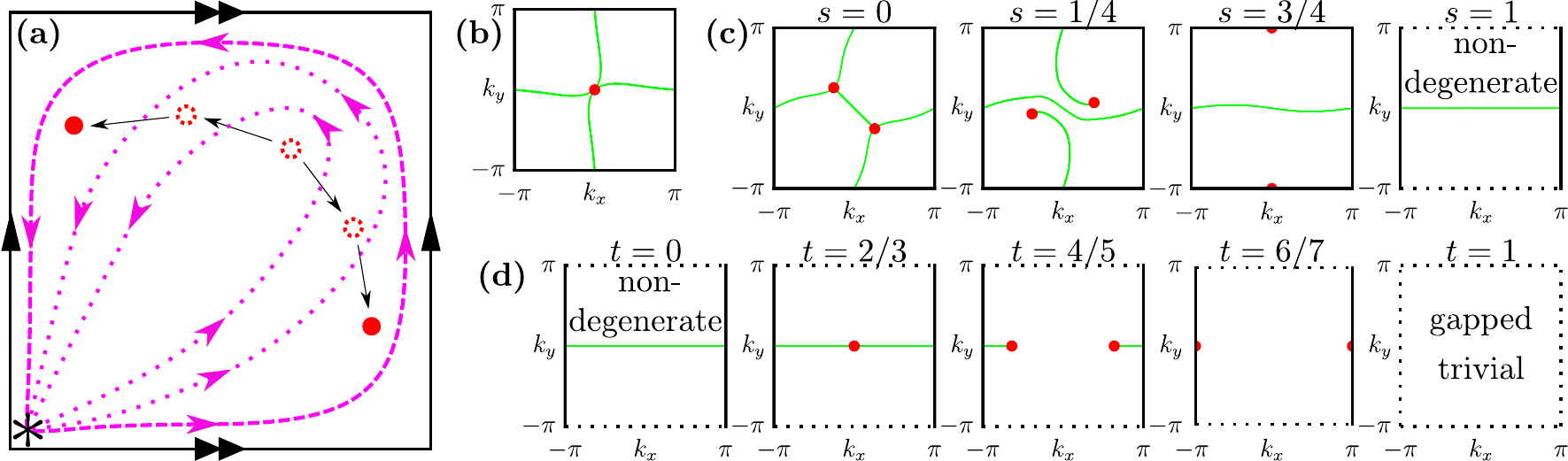}
	\caption{
    Non-Hermitian phase transition.
    \textbf{(a)} 
        To gap out the system, any contractible loop (fuchsia) in the BZ must carry a trivial braid. 
        The only way to change nontrivial braids (dashed) into trivial ones (dotted) continuously is to move an EP across the loop.
    \textbf{(b)} 
        Schematic representation of the spectrum of Eq.~\eqref{eq:tune-to-single} at \(\delta=1\). 
        The single nontrivial EP3 is marked in red, and the green lines illustrate the real-gap closings which form nontrivial loops around the BZ.
    \textbf{(c)} 
        Schematic representation of the spectrum for the interpolation Eq.~\eqref{eq:interpolation-to-gapped} between the unpaired model hosting unpaired EPs and a gapped phase.
        Initially the BZ boundaries (solid black lines) carry braid charges \(\sigma_1\) (\(\sigma_2\)) in the \(k_x\) (\(k_y\)) direction.
        At \(s=3/4\) the two EPs merge at the boundary, breaking the \(k_x\) braid charge into a trivial braid; the corresponding boundary is then marked with dotted lines.
        For \(s>3/4\), all braids along $k_y=\textrm{const}$ are trivial, while loops along $k_x=\text{const}$ correspond to braids of type \(\sigma_2\).
        The boundary braid in that regime is then $\sigma_2\sigma_2^{-1}=\mathds 1$.
    \textbf{(d)}
        Same as (c) for interpolation Eq.~\eqref{eq:interpolation-to-trivial} between nontrivial and trivial gapped phases for increasing interpolation parameter \(t\).
        A single twofold degeneracy is created at \(t=2/3\) and subsequently splits into two EP2s.
        These merge at the boundary for \(t=6/7\).
        For \(t>6/7\) the system is topologically trivial and gapped; all loops in this system correspond to the trivial braid.
        Note that the last panel in (c) and the first panel in (d)
        describe the same system.
        }
	\label{fig:cartoon}
\end{figure*}

Here, challenging the intuitive notion of fermion doubling, we reveal the possibility of topologically stable gapless phases in periodic systems that can host \emph{unpaired} EPs, i.e., nodal points that are not gapped out when fusing.
We showcase this in a tight-binding model hosting only two EPs in its BZ whose topological charges do not cancel. 
Instead of annihilating, these unpaired EP2s fuse to a single nontrivial threefold degeneracy (EP3).
We find that these NH topological band structures are enabled by noncommuting braids of the complex energy levels along noncontractible loops in the BZ, which remove the need for the EPs' topological charges to pair up and cancel (see Fig.~\ref{fig_sum} for an illustration).
The real part of the spectral gap closes along so-called Fermi arcs, which conventionally form open lines in the BZ that terminate in degeneracies. 
In the phases outlined here they can instead form closed and even noncontractible loops in the BZ. 
The only way to adiabatically gap out the nodal points is to move them along nontrivial loops in the BZ before annihilation (see Fig.~\ref{fig:cartoon}). 
Transitions between these NH phases are thus topological, and correspond to braiding of nodal points in the BZ.

\emph{Admissible EPs by the braid group.} 
In the vicinity of EPs, the eigenvalues are generally not single valued.
This is manifest in the braiding of the complex energies when encircling an EP, and endows these degeneracies with a topological charge \cite{PhysRevB.103.155129,PhysRevB.101.205417}. 
An illustrative example is an EP2 with dispersion $E=\pm\sqrt{k_x+ik_y}$, where two eigenvectors and eigenvalues coalesce. 
The two energy levels are swapped counterclockwise along a counterclockwise path around the EP. 
General \(N\)-fold degeneracies carry braid invariants with \(N\) strands. The corresponding braid group $B_N$ is generated by elementary braids $\sigma_j$ ($1\le j\le N-1$) that swap $E_j$ and $E_{j+1}$ counterclockwise in the complex plane. 
They satisfy the relations \cite{kassel2008braid}
\begin{equation}
\sigma_{j+1}\sigma_j\sigma_{j+1}=\sigma_j\sigma_{j+1}\sigma_j,\ \sigma_i\sigma_{j}=\sigma_{j}\sigma_{i} \,(|i-j|>1).\label{eq_hgprule}
\end{equation}
In contrast to standard Hermitian degeneracies such as Weyl points \cite{PhysRevLett.121.106402}, EPs are associated with non-Abelian groups, similar to topological defects in biaxial nematics \cite{RevModPhys.51.591,RevModPhys.84.497} or multigap phases \cite{PhysRevX.9.021013,wu2019non,bouhon2020non,jiang2021experimental,slager2022floquet}. This comes from the non-Abelian fundamental group $\pi_1$ of the space $\mathcal W_N$ formed by gapped non-Hermitian Hamiltonians \cite{PhysRevB.103.155129,PhysRevB.101.205417,suppm}.
The combination or split of EPs can be described through the standard group action \cite{RevModPhys.51.591}. 
As it is non-Abelian, one needs to be careful in fixing a base point $\ast$ when doing the group product. 
We call an EP trivial when a small loop around it corresponds to a trivial braid of eigenvalues, and nontrivial otherwise. 
This terminology is motivated by the fact that only trivial EPs can be gapped out by small perturbations.

We denote the homotopy class of each loop \(\gamma_j\) around the EP $j$ with a common base point $\ast$ as $[\gamma_j]\in B_N$. 
According to the topology of a torus, the total EPs in the BZ satisfy a non-Abelian sum rule \cite{PhysRevB.106.L161401,PhysRevResearch.4.L022064} [illustrated in Fig.~\ref{fig:ham-d-3D-spectra}(a)],
\begin{equation}
    [\gamma_1]\cdot[\gamma_2]\cdots[\gamma_n]=aba^{-1}b^{-1}\equiv [a,b],\label{eq_nabdoubling}
\end{equation}
where $[a,b]$ is the group commutator of two elements $a,b$ in $B_N$. 
They represent how the energy is braided along the meridian and the longitude of the torus (see Fig.~\ref{fig_sum}). 
As a result of this sum rule, models with a nontrivial commutator \([a,b]\) must contain degeneracies; they cannot be gapped.

As the group $B_N$ is non-Abelian and not free,
the sum rule can sometimes be hard to apply in practice. 
We can extract an easier necessary condition for EPs by taking the Abelianization $\pi_1/[\pi_1,\pi_1]$, which is the first homology group $\mathbf H_1(\mathcal W_N)$ \cite{bott1982differential}. 
Since all generators $\sigma_j$ are conjugate to each other, the Abelianization of the braid group is $\mathbb Z$ \cite{kassel2008braid}. 
The Abelianized element $ [\gamma]_A$ can be intuitively understood as the number of band permutations along a braid (the total number of $\sigma_j$'s in a braid). 
An $n$th root EP
\cite{hodaei2017enhanced,PhysRevLett.127.186601,Flore2022} has $[\gamma]_A=n-1$. 
The right-hand side of Eq.~\eqref{eq_nabdoubling} is the zero element $0$ in the homology group. 
The Abelianized sum rule is
\begin{equation}
    \sum_j[\gamma_j]_A=0,\label{eq_abdthe}
\end{equation}
where the sum represents the usual group product operation for Abelian groups. 
Equation \eqref{eq_abdthe} asserts that we cannot have a single EP with $n$th root dispersion in a BZ. 
Its Abelianized charge must be compensated, 
e.g., by a conjugate EP that has $[\gamma]_A=-(n-1)$.

However, if the net number of elementary braids around a set of EPs is zero, they satisfy the Abelianized sum rule and can occur in a BZ.
Depending on whether \(a\) and \(b\) commute, their topological charges can nevertheless be nontrivial and need not cancel out.
This is a significant difference from Abelian degeneracies, where periodic boundary conditions necessitate a trivial bulk. 
In the following parts, we explicitly show the existence of nontrivial unpaired EPs. 
In the Supplemental Material we further present tight-binding models that feature topologically trivial single EPs of arbitrary order in their BZ \cite{suppm}. 
They can be created locally and under perturbation either split into nontrivial paired EPs or get gapped out.

\emph{Unpaired EPs.} 
EPs without partners compensating their braid charge can only exist for models with three or more bands, since the group $B_2$ for two-band systems is Abelian, rendering eqs.~\eqref{eq_nabdoubling} and \eqref{eq_abdthe} identical. 
In the case of three bands, all braids are generated by $\sigma_{1}, \sigma_{2}$ and the relation \eqref{eq_hgprule}. 
Those that can be written as a commutator contain equal numbers of crossings \(\sigma_i\) and inverse crossings \(\sigma_j^{-1}\). 
A simple nontrivial such braid is $B= \sigma^{-1}_2\sigma_1$. 
By relation~\eqref{eq_hgprule}, it can be written as a commutator,
\begin{align}
    [\sigma_1,\sigma_2] 
    &= 
    \sigma_1\sigma_2\sigma_1^{-1}\sigma_2^{-1}
   =\sigma_2^{-1} (\sigma_2 \sigma_1\sigma_2)\sigma_1^{-1}\sigma_2^{-1}\nonumber
    \\&=\sigma_2^{-1} (\sigma_1 \sigma_2\sigma_1)\sigma_1^{-1}\sigma_2^{-1}=
    \sigma_2^{-1}\sigma_1\label{eq:braid-word-short}.
\end{align}

A simple model Hamiltonian carrying these braids is
\begin{equation}
    H_B =
    \mqty( 1
     -e^{i k_x} & 
    -i (1+e^{i k_x}) &
    0
    \\
    -i (1+e^{i k_x}) &
    e^{i k_x}-e^{i k_y} &
    -i (1+e^{i k_y})
    \\
    0 & 
    -i (1+e^{i k_y} )&
    -1 +e^{i k_y}
    ),
    \label{eq:simple-boundary-model}
\end{equation}
for \(k_x,k_y\) periodic, parametrized on \([-\pi,\pi]\).
It represents the momentum-space Bloch Hamiltonian of a real space model with nearest-neighbor hoppings on a two-dimensional square lattice, where each unit cell contains three orbitals. 
The spectrum along \((k_x,-\pi)\) corresponds to $a=\sigma_1$ , and to $b=\sigma_2$ along \((\pi,k_y)\). 
As shown in Eq.~\eqref{eq:braid-word-short}, these boundary braids do not commute.
The total BZ boundary braid, along the composition of the two great circles [Fig.~\ref{fig_sum}(b)], is the nontrivial braid \( \sigma_1\sigma_2\sigma_1^{-1}\sigma_2^{-1}\).
We show the spectrum on the BZ boundary in Fig.~\ref{fig:ham-d-3D-spectra} (b).
In the supplemental material \cite{suppm}, we extend the constructive method of Ref.~\cite{BodeDennis2016} to show that there is always a way to construct a given braid within a BZ-continuous model.

The entire spectrum of $H_B$ is shown in Fig.~\ref{fig:ham-d-3D-spectra} (c). 
It hosts two EP2s (which occur generically in two dimensions) that carry braid charges \(\sigma_2^{-1}\) and \(\sigma_1\), compensating the boundary braid according to the sum rule~\eqref{eq_nabdoubling} or \eqref{eq:braid-word-short} explicitly. 
While they have opposite Abelianization, their braid charges are not mutual inverses: These EPs are \emph{unpaired}.

To illustrate this further, we tune this Hamiltonian to 
\begin{equation}
    H^{\text{EP3}}_\delta = H_B + \delta\, 2\sqrt{2}\operatorname{diag}(1,0,-1),\quad 0\leq\delta\leq 1,
    \label{eq:tune-to-single}
\end{equation}
which leaves the boundary braids unchanged but merges the EP2s as \(\delta\to1\).
We show the corresponding spectrum in Fig.~\ref{fig:ham-d-3D-spectra}(d).
As the EP2s are unpaired, they cannot annihilate, but fuse to a \emph{single} EP3 at \(k_x=k_y=0\).

While fine tuned, this EP3 is significantly distinct from the simple third-root EPs commonly implemented in topologically trivial spaces \cite{PhysRevLett.126.086401,PhysRevLett.127.186601,PhysRevLett.127.186602}.
First, it can exist alone under the BZ periodic boundary condition.
Second, its dispersion consists of the roots of a third-order polynomial which forbids a simple expression.
Third,  it lies at the intersection of two Fermi arcs, along which one of the real gaps $\operatorname{Re}(E_i-E_j)$ closes. 
These Fermi arcs extend through the BZ, roughly following \((k_x, 0)\), and \((0, k_y)\). 
In particular they form extended noncontractible loops, unlike the common line-segment Fermi arcs which are known to appear in topologically trivial models.

\emph{Topological robustness of the gapless phase.} 
The topological robustness of the EP phase arises from the nontrivial braids along contractible loops on the BZ torus, especially the loop successively traveling as $aba^{-1}b^{-1}$. 
We show first that all contractible loops in a gapped area must carry trivial braids:
We take a deformation retraction \(\gamma_s\) ($0\le s\le 1$) between some loop \(\gamma_0\) in the BZ and a constant loop \(\gamma_1\) (a point). 
If the Hamiltonian is gapped throughout this deformation retraction, then there is a well-defined braid of eigenvalues \([\gamma_s]\) corresponding to each \(\gamma_s\), defining a continuous deformation of braids. 
Such continuous deformation without intersecting strands does not affect the braid group element, which means \([\gamma_0] = [\gamma_1]\). As the constant loop carries the trivial braid this concludes the proof.
This implies conversely that if any contractible loop in the BZ carries a nontrivial braid, the system must be gapless.

This argument implies robustness of the gaplessness under perturbation.
In the BZ of \(H_B\), all contractible loops enclosing either or both of the EPs carry nontrivial braids. 
The only way to break the braid along a loop is to move EPs across it [Fig.~\ref{fig:cartoon}(a)]. 
When perturbing $H_B$, the braids along loops far away from the EP are unaffected. 
They are of the same type as labeled by the great circles $aba^{-1}b^{-1}$. 
So the system remains gapless and contains multiple EP2s or a single EP3, whose braids combine to $\sigma^{-1}_2\sigma_1$. 

It is instructive to consider how the degeneracies evolve when we transition from the gapless phase to a gapped phase. We interpolate between Eq.~\eqref{eq:simple-boundary-model} and a model $H_B(\pi,k_y)$ without \(k_x\) dependence,
\begin{equation}
    H_s(k_x,k_y)
    = 
    (1-s)H_B(k_x,k_y) + sH_B(\pi,k_y).
    \label{eq:interpolation-to-gapped}
\end{equation}
Physically this corresponds to decoupling the unit cells in the \(x\) direction.
The results are summarized in Fig.~\ref{fig:cartoon}(c). 
As $s$ increases, the two EPs move away from each other through the BZ towards $(0,\pm\pi)$. Their combined trajectory traces over the meridian along a large nontrivial loop involving the boundary.
Finally,  for \(s=3/4\) they fuse at \((0,\pi)\). 
Further increasing $s$ annihilates the two EPs and we obtain a gapped phase.

We may also understand this process and the robustness of the phase via the Fermi arcs.
The initial model hosts two topologically nontrivial Fermi arcs, one along each great circle.
These arcs can only end in degeneracies, and they change continuously under continuous deformation of \(H_B\).
By removing dependence on \(k_x\), we remove the Fermi arc that follows \(k_y\), correspondingly, two degeneracies must follow the meridian around the torus.

The $s=1$ phase, while nondegenerate, is still a topologically nontrivial phase. 
One of the three bands is effectively decoupled from the others, but 
the other two are braided in \(k_y\) direction.
They share a corresponding extended Fermi arc of $\operatorname{Re}(E_2 - E_3)$ closing along \((k_x,0)\).
The spectrum along the BZ boundary corresponds to $(a,b)=(\mathds{1},\sigma_2)$.
These two braid elements commute and the combined braid along the boundary is trivial, consistent with previous classification \cite{PhysRevB.103.155129,PhysRevB.101.205417}.
We also present how this topologically nontrivial gapped phase can be switched into a fully topologically trivial phase $H_0=[2+\cos(k_x)]\mathrm{diag}(1,0,-1)$. We choose the following interpolation,
\begin{equation}
    H_t=
    (1-t)
    H_B(\pi,k_y)
    + t H_0.
    \label{eq:interpolation-to-trivial}
\end{equation}
The evolution of the spectrum is shown in Fig.~\ref{fig:cartoon} (d). 
At \(t=2/3\), a new degeneracy is created at \(\mathbf{k}=(0,0)\). It splits into a pair of EP2s for increasing \(t\), which then separate further in the $k_x$ direction. 
Their trajectory traces over a complete longitude, until they merge and annihilate into the trivial gapped phase at \(t=6/7\).

The above series of figures showcases how the topology of EP-annihilation trajectories corresponds to different phase transitions. 
In the $H_s$ interpolation~\eqref{eq:interpolation-to-gapped}, the EP2s travel over the meridian to cross all contractible loops enclosing the EP3 and break braid \(a\). 
In the Supplemental Material \cite{suppm}, we prove that if the two EP2s trace out a simple loop, it must be in the noncontractible homotopy class, based on the universal covering map $\mathbb R^2\to T^2$ and the Jordan-Schoenflies theorem \cite{munkres2000topology,munkres2018elements}. In the $H_t$ interpolation~\eqref{eq:interpolation-to-trivial}, the newly created EPs must cross all meridians in order to render the $b$ braid trivial.

\emph{Conclusion.} 
We have constructed non-Hermitian phases with unpaired exceptional degeneracies and explained their topological robustness, illustrated in a short-range tight-binding lattice model. 

The non-Abelian nature of three or more complex energy bands allows for overall nontrivial topological charges and even non-Abelian monopoles under periodic boundary conditions in reciprocal space.
In nontrivial phases these charges are unpaired and cannot be annihilated locally by perturbations. 
Instead they must travel nontrivially around the BZ torus in order to be gapped out.
This endows the gapless phases presented here with topological robustness, as the movement of degeneracies generally is continuous when deforming a model.

Conversely, topologically inequivalent ways of moving EPs around the BZ lead to distinct gapped and gapless phases. 
The movement of EPs thus encodes the topological information of the system.
This differs from the situation on a plane \cite{guo2022exceptional}, where at least three EPs are required to exhibit non-Abelian features.

Under some tuning, our model exhibits an isolated nontrivial higher-order degeneracy with dispersion beyond the \(n\)th-root behaviours constructed previously in tight-binding models.
This EP3 is accompanied by extensive Fermi arcs that form noncontractible loops, instead of open line segments. 
These topological Fermi arcs do not have counterparts in Hermitian Weyl nodes, and are purely a feature of non-Hermitian systems.

\emph{Note added}: After the first version of our work, two experimental works  \cite{zhangObservationAcousticNonHermitian2023,zhang2022experimental} realized non-Hermitian multiband systems with nontrivial energy braids in acoustic metamaterials, which corroborates the direct experimental relevance of our work.

\acknowledgments{{\em Acknowledgments.} We thank Lukas R\o{}dland, Marcus St\r alhammar and Zhi Li for stimulating discussions. J.L.K.K., K.Y. and E.J.B. were supported by 
the Swedish Research Council (VR, Grant No. 2018-00313), 
the Wallenberg Academy Fellows program (2018.0460) and 
the project Dynamic Quantum Matter (2019.0068) of the Knut and Alice Wallenberg Foundation, 
as well as the  G\"oran Gustafsson Foundation for Research in Natural Sciences and Medicine. 
K.Y. is also supported by the ANR-DFG project (TWISTGRAPH).}

\bibliography{singleEP}

\clearpage

\renewcommand{\theequation}{S\arabic{equation}}
\setcounter{equation}{0}
\renewcommand{\thefigure}{S\arabic{figure}}
\setcounter{figure}{0}
\renewcommand{\thetable}{S\arabic{table}}
\setcounter{table}{0}
\begin{widetext}
\section{Supplemental Material}

\subsection{Braid groups and exceptional points}

The tour around the EP gives rise to a map, denoted as $\gamma$, from a circle $S^1$ to the space spanned by gapped NH Hamiltonians $\mathcal W_N=[\mathrm{Conf}_N(\mathbb C)\times GL_N(\mathbb C)/(\mathbb C^\times)^N]/S_N $ \cite{PhysRevB.103.155129,PhysRevB.101.205417}. 
The group $GL_N(\mathbb C)$ represents the general linear transformation to the eigenbasis, from which we quotient out the transformations that give identical diagonalization. 
These are scaling transformations $\mathrm{diag}(z_1,z_2,\dots, z_N)$ with $z_j\ne 0$, represented by $(\mathbb C^\times)^N=(\mathbb C\setminus\{0\})^N$. 
The configuration space $\mathrm{Conf}_N(\mathbb C)$ corresponds to the energy spectrum, which is formed by lists of $N$ distinct complex numbers $\{E_1,E_2\dots,E_N\}$. 
The group $S_N$ consists of joint permutations of eigenvalues and eigenvectors. 
The topology of this map can be described by the homotopy group $\pi_1[\mathcal W_N]=B_N$ \cite{PhysRevB.103.155129,PhysRevB.101.205417}, which is the braid group on \(N\) strands.

\subsection{Base-point homotopy vs free homotopy descriptions}

The topology of a point degeneracy or defect in two dimensions is described by how the gapped Hamiltonian or the order parameter evolves along a loop around it. There are two different but mutually complementary ways of describing it, the base-point homotopy theory and the free homotopy theory. The former has the advantage of describing operations on the degeneracy (defects) through the powerful group theory, while it needs to artificially assign a base point somewhere. The latter is independent of such choice and more natural in intuition. However, the free homotopy theory does not have a group structure and is not appropriate when studying what kind of defects or degeneracy can merge or split. This is the reason why we use the base-point homotopy in the main text. In this section, we introduce its connection to the freely homotopic classes.

The base-point homotopy group $\pi_1(X,x_0)$ is defined with respect to all maps $\gamma$ from $S^1$ to some space $X$ with a fixed point in $S^1$ mapped to a fixed point $x_0\in X$. To make the language clearer, we identify $S^1$ as the unit interval $[0,1]$ with periodic boundary condition. The base-point homotopy group is to study all continuous maps $\gamma: [0,1]\to X$ satisfying $\gamma(0)=\gamma(1)=x_0$. The point $x_0$ is called the base point. The base point plays an important role in constructing a group structure, such that different loops can be composed from the common base point. The composition of two loops $\gamma_1\cdot\gamma_2$ is regarded as a loop that first travels from $x_0$ along $\gamma_1$ to $x_0$, and then do the second leg along $\gamma_2$. Two maps $\gamma_1$ and $\gamma_2$ starting at $x_0$ are homotopic if they can be interpolated continuously by a series of loops from $x_0$ (Fig.~\ref{fig_EPpro}). The homotopic class of $\gamma$ is denoted by $[\gamma]$.

For a path-connected space, one can show that the homotopy groups $\pi_1(X,x_0)$ at different base points only differ by an isomorphism. Taking two base point $x_0$ and $x'_0$, as the space is path-connected, we can find a path $\gamma_0$ with joints them. A loop $\gamma$ starting from $x_0$ can be extended to a loop from $x'_0$ by $\gamma_0\cdot\gamma\cdot\gamma^{-1}_0$. So $\gamma_1\cdot\gamma_2$ is mapped to $\gamma_0\cdot\gamma_1\cdot\gamma^{-1}_0\cdot\gamma_0\cdot\gamma_2\cdot\gamma^{-1}_0$. Notice that the loop $\gamma^{-1}_0\cdot\gamma_0$ is homotopic to a constant loop. So we have $[\gamma^{-1}_0]\cdot[\gamma_0]=\mathds{1}$. When taking the homotopic classes, this correspondence becomes $[\gamma_1]\cdot[\gamma_2]\to [\gamma_0]\cdot[\gamma_1]\cdot[\gamma_2]\cdot[\gamma^{-1}_0]$, preserving the group product. The inverse homomorphism from $\pi_1(X,x'_0)$ to $\pi_1(X,x_0)$ is given by the action of $\gamma_0^{-1}$. Thus $\pi_1(X,x'_0)$ and $\pi_1(X,x_0)$ are isomorphic. Under such identifications, one may not write the base point out explicitly and use $\pi_1(X)$ to denote the abstract structure of the base-point homotopy group.

Now we introduce the freely homotopic class $[S^1,X]$. In this definition, two loops $\gamma_1$ and $\gamma_2$ belong to the same class if they can be interpolated continuously by a series of loops in $X$, without requiring a fixed starting point (see Fig.~\ref{fig_EPpro}). Compared to the base-point homotopy, the free homotopy allow more ways of identifying loops. And it characterizes the defects in a more invariant way as there is no artificial choice of references. However, due to the lack of a base point, there is no group structure in $[S^1,X]$. It is not a good tool when talking about the composition or separation of degeneracy. 

In order to see the connection between the two homotopic classes, one can consider a loop $\gamma'$ from $x_0$ to itself. Its action on $\pi_1(X,x_0)$ is an automorphism of the base-point homotopy: $[\gamma']\cdot\pi_1(X,x_0)\cdot[\gamma^{\prime-1}]$. This operation can be considered as a free homotopy which moves the base point $x_0$ along $\gamma'$ \cite{bott1982differential}. So this is a map from the base-point homotopic classes to the free homotopic classes. One can further prove that this is indeed a bijection map. The freely homotopy classes can be identified by \cite{bott1982differential}
\begin{equation}
    \pi_1(X,x_0)/\pi_1(X,x_0)\simeq [S^1,X],\label{eq_bphfh}
\end{equation}
where the quotient should be understood as the conjugate action introduced.

In the base-point homotopy language, the EP is characterized by the braid group element $[\gamma]$. From the above discussion, if we choose a different base point but the same loop, the corresponding braid group element becomes $[\gamma_0]\cdot[\gamma]\cdot[\gamma_0^{-1}]$. So in the base-point homotopy description, the correspondence between braid elements and the topological types of EP is not unique. Braid elements conjugated to each other can be equivalent EPs. This ambiguity is resolved in the freely homotopic description. For the case of non-Hermitian EPs, the classes in $ [S^1,\mathcal W_N]$ can be regarded as gluing the start configuration and end configuration of a braid together. It is a actually a knot embedded in a solid torus \cite{PhysRevB.103.155129,PhysRevLett.126.010401,PhysRevResearch.4.L022064}. 

\begin{figure}
    \centering
    \includegraphics[width=0.8\linewidth]{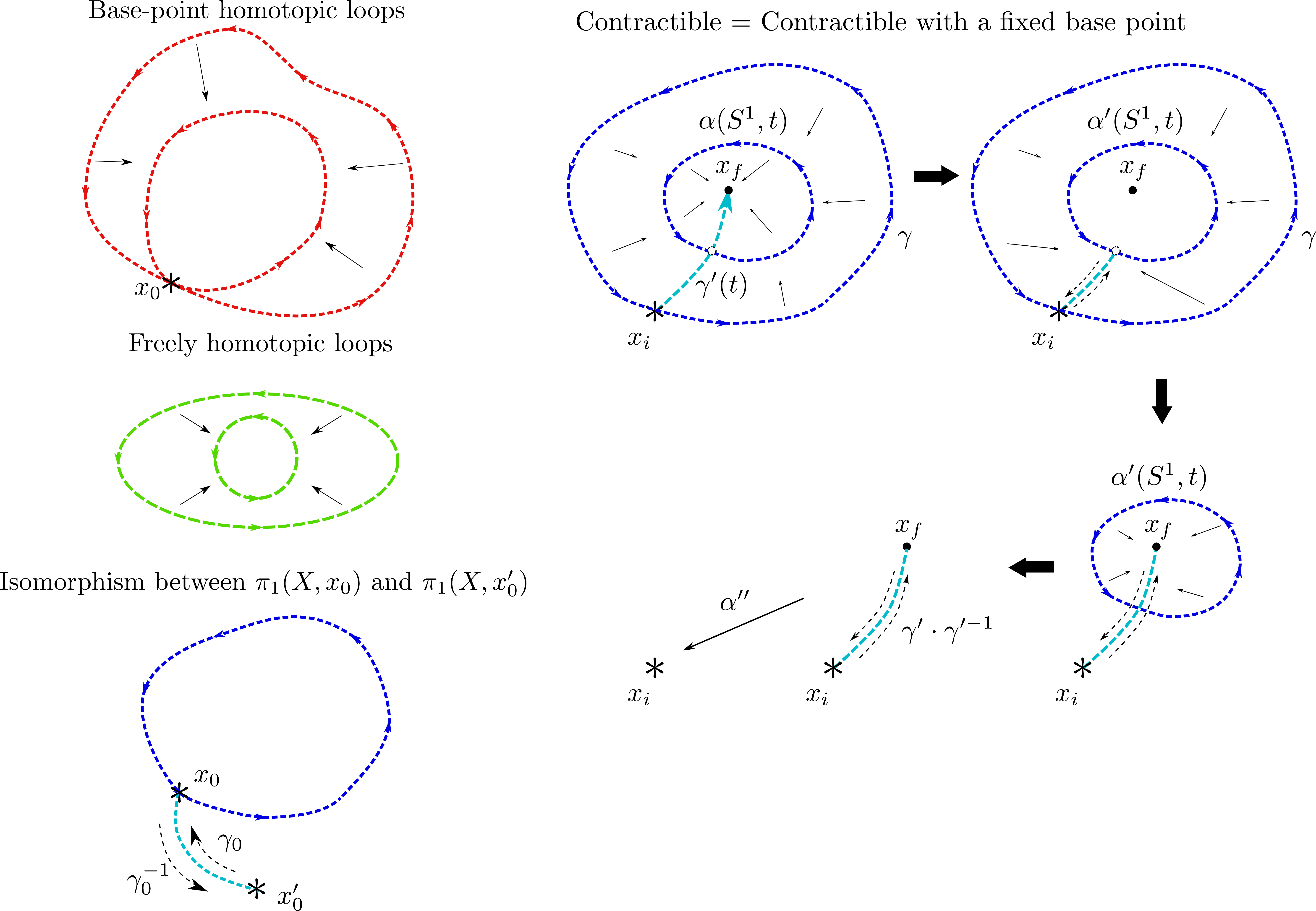}
    \caption{Left: the definition of base-point homotopic loops and freely homotopic loops and the isomorphism between the homotopy groups with respect to different base points. Right: When a loop is contractible to a point $x_f$, it is also contractible when a base point is fixed. This can be shown be combining the contraction with a path connecting the base point to the point $x_f$.}
    \label{fig_EPpro}
\end{figure}

To conclude this section, we prove the following statement used in the main text:
\begin{itemize}
    \item \emph{If a loop $\gamma:S^1\to X$ is contractible, it is also contractible when a base point is fixed. That is, a trivial element in the freely homotopic class $[S^1,X]$ is also trivial in the base-point homotopic class $\pi_1(X)$.}
\end{itemize}

Although this statement may be observed immediately from the relation \eqref{eq_bphfh}, we here give a detail construction for the equivalence (also Fig.~\ref{fig_EPpro}), in order to complement the discussion of braid protection in the main text. As before, we treat $S^1$ as the unit interval $[0,1]$ with periodic boundary condition. The loop is written as map $\gamma:[0,1]\to X$. The contraction is a map $\alpha: S^1\times[0,1]\to X$ with $\alpha(s,0)=\gamma(s), \alpha(s,1)=x_f$.  Without loss of generality, we take the point $x_i=\gamma(0)$ as the base point. We show that we can find a map $\beta:S^1\times[0,1]\to X$ such that $\beta(s,0)=\gamma(s),\beta(s,1)=x_i$ and $\beta(0,t)=\beta(1,t)=x_i$. 

During the contraction $\alpha$, the point $x_i$ actually trace out a path from $x_i$ to $x_f$. We use $\gamma'$ to represent this path. For each $t$, we notice that $\alpha(S^1,t)$ is a loop starting from $\gamma'(t)$. We label the path from $\gamma'(0)$ to $\gamma'(t)$ as $\gamma_t$. Then $\gamma_t\cdot \alpha(S^1,s)\cdot\gamma_t^{-1}$ is a loop starting from $x_0$, traveling to $\gamma'(t)$, doing a loop along $\alpha(S^1,t)$ and traveling back to $x_0$. By doing this tour for every $t$, we construct a series loops $\alpha':S^1\times [0,1]\to X$ with $\alpha'(s,0)=\gamma(s), \alpha'(s,1)=\gamma'\cdot\gamma^{\prime-1}(s)$ and $\alpha'(0,t)=\alpha'(1,t)=x_i$. This is a base-point homotopy between $\gamma$ and the composition path $\gamma'\cdot\gamma^{\prime-1}$. The latter obviously is homotopic to a constant map at $x_0$ via some map $\alpha^{\prime\prime}$. So the desired base-point homotopy from $\gamma$ to a constant map is given by $\beta=\alpha'\cdot\alpha^{\prime\prime}$.

\subsection{Trivial single EPs of any order}

We first present single EP2s corresponding to trivial eigenvalue braids. 
Consider
\begin{equation}
H=\begin{pmatrix}
           0 & 1 \\
       f(\mathbf k) & 0   
         \end{pmatrix}
\end{equation}
which has eigenvalues $E=\pm \sqrt{f(\mathbf k)}$ and an exceptional point at $f(\mathbf k)=0$. 
With $ f(\mathbf k)=2-e^{i k_x}-e^{i k_y}$, we have a single EP at $\mathbf k=0$ in a periodic two-dimensional model. 
Near the EP, we have $f(\mathbf k)\simeq -i(k_x+k_y)+|k|^2/2$, thus the dispersion is square root in generic directions, whereas it is linear only along the high-symmetry line $k_x=-k_y$ along which the linear terms cancel.

This degeneracy is however only stable to particular perturbations. 
Any perturbation around the EP can be represented by replacing $f(\mathbf k)$ by $f(\mathbf k)+\epsilon$ \cite{jiang2020nonreciprocal,Flore2022}. If we want to keep the system gapless, there should exist a solution to $\epsilon-i(k_x+k_y)+|k|^2/2=0$. So we must have $\textrm{Re }\epsilon=-|k|^2/2$ and $\textrm{Im }\epsilon=(k_x+k_y)$. The second equation can always be satisfied while the first equation imposes restrictions on $\epsilon$.
The EP is therefore gapped out if $\textrm{Re }\epsilon>-|\textrm{Im}\epsilon|^2$, 
while splitting into pairs of nontrivial EP2s for $\textrm{Re }\epsilon<-|\textrm{Im}\epsilon|^2$.

We show how trivial EPs can be generalized to more situations. For the $2\times 2$ matrices with $1$ and $f(\mathbf k)$ on the off diagonal, there is a single EP at $\mathbf k=0$ with the choice in $d$ dimensions 
\begin{equation} 
f(\mathbf k)=d-\sum_{j=1}^d e^{i k_j}.\label{f}
\end{equation} 
This gives a single EP2 at ($\mathbf k=0$) in a periodic model in any dimension $d$. Again the dispersion is square-root in generic directions whereas it is linear along high-symmetry lines along which the linear terms cancel, i.e.  $\sum_{\{j\}}k_j=0$ summed over some (sub)set of momenta.  

The scenario above for second order EPs generalize directly to any order. To this end consider the $n \times n$ matrix
\begin{equation}
H_n=\begin{pmatrix}
    0 & 1 & & & & & & & \\
    0 & 0 & 1 & & & & & & \\
    & 0 & 0 & 1 & & & & & \\
    & & 0 & 0 &  & & & & \\
    & & & & & \ddots &  & & \\
    & & & & & & 0 & 1 &   \\        
    & &   & & & & 0 & 0 & 1 \\
    f(\mathbf k)& &  & & & & &  0 & 0   
\end{pmatrix}, 
\end{equation}
which has eigenvalues fulfilling $(-E)^n=f(\mathbf k)$ hence $E\propto \sqrt[n]{f(\mathbf k)}$. At $f(\mathbf k)=0$ there is an $n$-th order EP, i.e. an EP$n$. Here we note that with the choice of $f$ in Eq. (\ref{f}) the dispersion is generically $\sim k_1^{1/n}$ and in high symmetry directions maximally $\sim k_1^{2/n}$, hence it has the telltale EP feature of a sublinear dispersion in all directions. Of course it also has a single right (and left) eigenvector.

\subsection{Construction of unpaired degeneracies on a torus from given braids}

We recapitulate the construction of a Hamiltonian that has a point degeneracy with some specific braid \(B\) as a topological invariant, as presented in Ref.s~\cite{PhysRevLett.126.010401,PhysRevResearch.4.L022064}.
This builds on the mathematical construction in Ref.~\cite{BodeDennis2016} of a polynomial with a given braid as its root set.
We show that a continuous extension to the torus is possible whenever the braid \(B\) is a commutator, albeit this may require infinite hopping range, not as simple as the model in the main text.

The construction starts from building a Hamiltonian that gives the desired braid in the vicinity of the EP. We begins with the following ansatz
\begin{equation}
    H_c(k)= \mqty(
        -c_{M-1}(k) & & \cdots& &-c_0(k)\\
        1 & 0 & \cdots & \cdots & 0\\
        0 & \ddots & \ddots &&\vdots\\
        \vdots & \ddots & \ddots&\ddots&\vdots\\
        0 & \cdots &0 & 1 & 0\\,
    )
\end{equation}
which has as its eigenvalues the roots \(\lambda(k)\) of 
\begin{equation}
    p(\lambda,k) = \lambda^M+\sum_{j=0}^{M-1} c_j(k)\lambda^j.
\end{equation}
We set $\mathbf k=0$ as the EP position. The goal is then converted to constructing a characteristic polynomial with the appropriate roots. 
This reduces further to finding a simple expression for the roots as complex functions \(Z_i(k)\), since then the polynomial
\begin{equation}
    p=\prod_{j=1}^M \left(\lambda - Z_j(k)\right)
\end{equation}
has the desired root set.

However, finding an expression for the roots is nontrivial, since they are multi-valued functions whenever the braid \(B\) considered is not pure.

Our construction proceeds as follows.
Following Ref.~\cite{BodeDennis2016}, we construct the polynomial \(p(k=\exp(i\phi))\) on a circle, by a truncation of the Fourier expansion of the braid diagram. 
We then extend it to a disk \(\set{z\in\mathbb{C}|\abs{z}<=1}\) with this circle on the boundary, such that the corresponding Hamiltonian is degenerate at a single point, following Ref.~\cite{PhysRevLett.126.010401}.
Finally we embed this disk in a torus, defining \(p\) on \(T^2\).

Assume without loss of generality that \(B\) consists of only one cyclic permutation cycle of length \(M\), i.e. following strand \(1\), we end at the starting point of strand \(2\), and so does this to strand $2$ and $3$, until some strand \(M\) that ends in the starting point of strand \(1\).
Models containing multiple such cycles can be obtained as tensor products of the one-cycle case or by multiplying their corresponding polynomials \(p\). 
Some care must be taken to avoid degeneracies between the constituent components.

The braid can then be represented by its braid diagram, which embeds into the complex numbers as \(M\) piecewise linear strands.
We choose here to represent the ordering of the \(M\) strands along the real line, and overcrossing, respectively undercrossing, of strands by the difference in imaginary parts at the points where the real parts cross.

Since the braid \(B\) consists of one \(M\)-cycle by assumption,
the traversal of all \(M\) strands in succession is periodic.
The braid can thus be represented by a single \(2 \pi\)-periodic function \(f\), that follows strand \(i\) on the \(i\)th interval of length \(2\pi/M\) in its domain.

This function \(f\) is chosen to be piecewise linear, and thus at least \(1\)-H\"older. Hence there is a triangular-polynomial approximation \(S_Nf\) that converges to it uniformly,
\begin{equation}
    \forall \phi\in[0,2\pi]:\ 
    \abs{\Delta_N(\phi)}
    =
    \abs{(S_Nf - f)(\phi)}
    \leq
    C_1 \ln(N)/N.
    \label{eq:unif-conv-f}
\end{equation}
We can therefore approximate the desired polynomial \(p\) using this \(S_Nf\), via
\begin{equation}
    p_N(\phi,z)=
    \prod_{s=1}^{M}
    \left(
        z- S_Nf\left(\frac{\phi}{M} + \frac{2\pi s}{M}\right)
    \right)
\end{equation}
This polynomial is well-defined, since the terms are simply permuted \(s \to s+1 \mod M\) as \(\phi \to \phi+2\pi\).

We need to take care that no two factors in this product are the same, to keep the resulting Hamiltonian nondegenerate.
This is the case if we either manually post-process \(S_N\) as is done in Ref.\cite{BodeDennis2016},
but choosing sufficiently large \(N\) is going to satisfy this automatically since any given braid has a finite minimal gap, and choosing large \(N\) keeps \(S_Nf\) close to \(f\) s.t. there are no added crossings due to the uniform convergence we show in the next subsection.

Even though the Fourier series is evaluated for \(\phi/M\), the authors show that approximations of this type result in polynomial dependence of \(p_N\)on \(\exp(\pm i\phi)\), i.e. \(p_N=p_N(\lambda,\exp(i\phi))\).
As shown in Ref.~\cite{PhysRevResearch.4.L022064}, this polynomial can be extended to the unit disk in the complex plane by replacing \(\exp(i\phi) = z/\abs{z}\), and possibly regularising to the polynomial
\begin{equation}
    \tilde{p}_N(\lambda,z) = \abs{z}^{g*M} p_N\left(\lambda/\abs{z}^{2g},z/\abs{z}\right)
\end{equation}
with some integer \(g>0\).
The extension to the disk is done for some sufficiently large but fixed \(N=N_0\).

Next, we show that this model can be continuously extended from the unit disk to the torus.
Note first that \(p_N\to p\) uniformly in \(\phi\), since
\begin{align}
    P_N(z,\phi) 
    &=
    \prod_{s=1}^{M}
    \left(
        z- S_N f\left(\frac{\phi}{M} + \frac{2\pi s}{M}\right)
    \right)
    \\
    &=
    \prod_{s=1}^{M}
    \Bigg(
        z-f(\ldots) + \underbrace{f(\ldots) -  S_N f\left(\frac{\phi}{M} + \frac{2\pi s}{M}\right)}_{=:\Delta_N^{[s]}\left(\phi\right)}
    \Bigg)
    \\&=
    P(z,\phi)+
    \sum_{m=0}^{M-1}\Delta_N^{[m]}(\phi)
    \prod_{s=m+1}^{M-1}
    \left(z-f\left(\frac{\phi}{M} + \frac{2\pi s}{M}\right)\right)
\end{align}
by partial expansion of the product. Thus we obtain the bound
\begin{equation}
    \abs{P_N-P}\leq
    M C_1 C_2^M \ln(N)/N
\end{equation}
with \(C_2 = \max_\phi\abs{z-f(\phi)}\) from the uniform convergence of \(S_N f\to f\) in Eq.~\eqref{eq:unif-conv-f}.

To extend the model to the torus, we simply restore the higher order terms \(S_N f, N>N_0\) to the triangular-polynomial approximation s.t. at the boundary of the chosen torus unit cell we have the model defined by \(f\), respectively \(p\).
This means that the eigenvalues of the Hamiltonian on this boundary are precisely the piecewise linear eigenvalues defined by the braid diagram. 
For a braid \(B\) that is a commutator, this has the required symmetry on the boundary, and is thus a well-defined map on the torus. 

The constructive procedure above guarantees a BZ-continuous model. However, in general it does not lead to a differentiable or analytic models in $\mathbf k$. This means infinite-range hopping terms may be required.

\subsection{Topological robustness of the EP3 or pair of EP2}

\emph{Annihilation condition}: According to the discussion in the main text, after the EP$3$ split into two EPs under perturbation, these two EPs cannot be directly brought together to be gapped out, since any contractible loop enclosing them carries a nontrivial braid. We denote the location of the EP3 as $x_0$ and a small disc containing it as $D_0$ (Fig.~\ref{fig_ep}). In order to break these braids protecting the degeneracy, the two EPs must trace over some large loop $C$ and cross all loops homotopic to $\partial D_0$ on $T^2-\{x_0\}$. After this they can be annihilated. 

\begin{figure}
    \centering
    \includegraphics[width=0.7\linewidth]{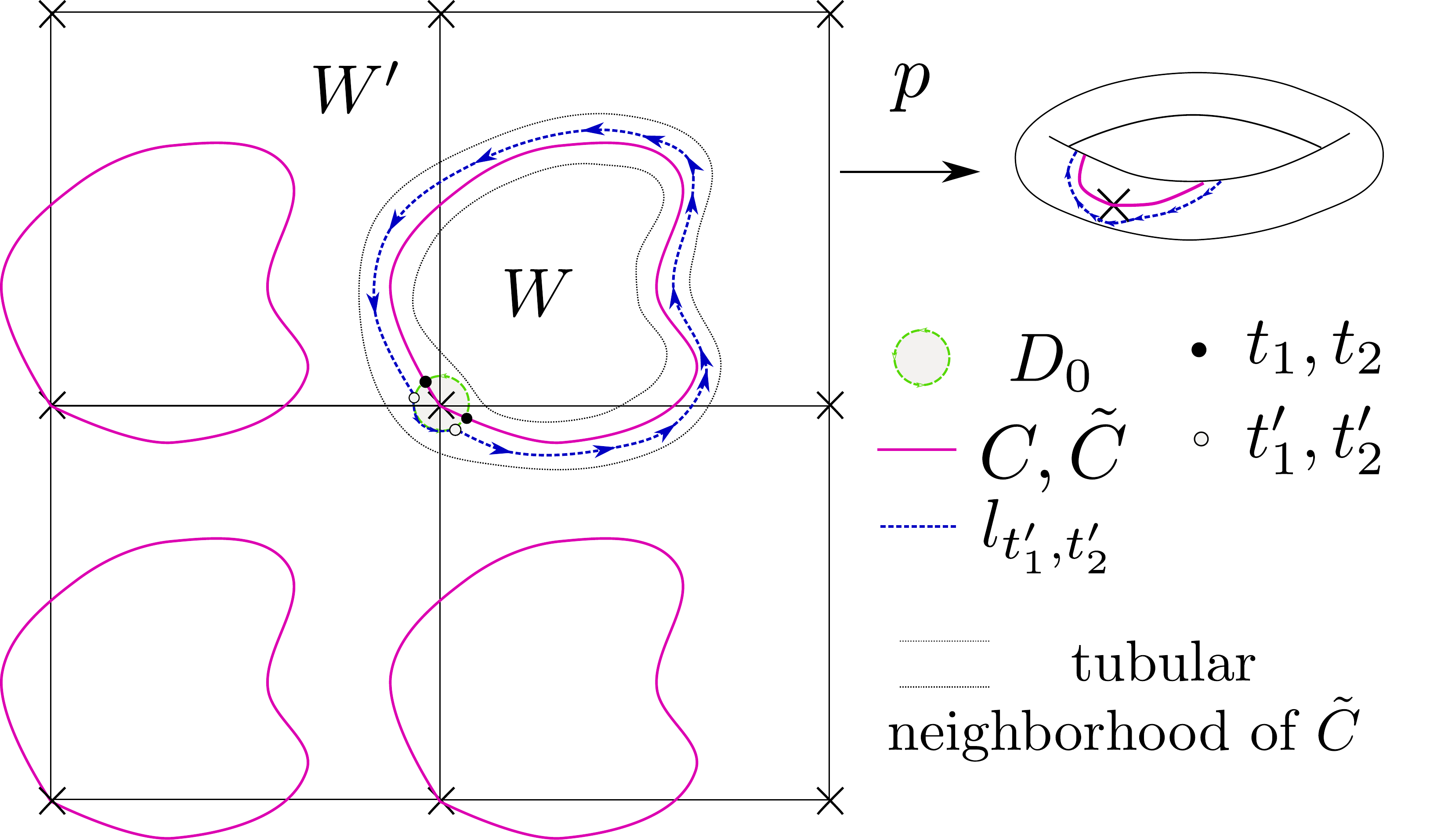}
    \caption{If the two EPs trace out a simple loop $C$ that is contractible on the torus, we can lift it to the covering space $\mathbb R^2$. There we can always construct a loop homotopic to the circle around the EP, while it does not intersect with the trajectory of EP, using the tubular neighborhood of $C$. So the nontrivial braid along it is not broken. Thus the loop $C$ must be some nontrivial loops on the torus.}
    \label{fig_ep}
\end{figure}

\emph{Topology of the EP trajectory}: In order to study the topological property of the curve $C$, we may assume that one EP is fixed and $C$ is traced over by the other EP. Then $C$ is regarded as a loop starting and ending at $x_0$. An immediate consequence is that $C$ must either intersect with all meridians or all longitudes. Otherwise we can take the nonintersected meridian and longitude as the boundary of the BZ. There is obviously a loop homotopic to $\partial D_0$ that does not cross $C$.

First, we prove that if $C$ is a simple loop that does not have any self-intersection, it must belong to the nontrivial classes of $\pi_1(T^2,x_0)$. Notice here we are interested in the image of $C$ instead of the map $C: S^1\to T^2$ itself. Only in the simple loop situation is the answer directly related to the homotopy group $\pi_1(T^2,x_0)$.

To make the question easier, we use the universal covering map $p: \mathbb R^2\to T^2$ to lift $C$ from the BZ to the two-dimensional plane. The advantage of lift is that we have a clear definition for the interior and the exterior of a simple loop. And we choose $p[(2m\pi,2n\pi)]=x_0$. We look for contradiction if $C$ is a contractible loop. If $C$ belongs to the trivial class of  $\pi_1(T^2,x_0)$, it has a unique lift $\tilde C$ starting and ending at $(0,0)$ \cite{munkres2000topology}, which is also a loop. Since $C$ does not intersect with itself, either does $\tilde C$. The preimages $p^{-1}(C)$ are made up of all translations of $\tilde C$. Moreover, the translations of $\tilde C$ do not intersect with $\tilde C$ since $C$ is a simple loop. So there is always a neighborhood of $\tilde C$ that contains no other preimages of $p^{-1}(C)$. According to the Jordan–Schoenflies theorem \cite{munkres2018elements}, $\tilde C$ separates $\mathbb R^2$ into two connected components  $W,W'$, with $\tilde C$ their common boundary. The enclosure of the bounded component (interior) $\bar W$ is actually homeomorphic to a closed disc. We can prove that $\bar W$ does not contain any points of $p^{-1}(x_0)$ except for $(0,0)$. If not so, $\bar W$ contains another $y\in p^{-1}(x_0)$, the lift $\tilde C'$ of $C$ starting from $y$ is a connected space on $\mathbb R^2-\tilde C$. So it must lie in $\bar W$. However this is not possible, since $\tilde C'$ is a translation of $\tilde C$. 

We assume that $\tilde C$ is an embedding of $S^1$ in $\mathbb R^2$. It has a tubular neighborhood $N$ that is diffeomorphic to its normal bundle \cite{bott1982differential}. This normal (line) bundle is trivial as it is orientable (from the orientability of $\tilde C$ and $\mathbb R^2$). So it can be taken as $\tilde C\times (-1,1)$ with $\tilde C\times 0$ identified with $\tilde C$. The connected component $\tilde C\times (0,1)$ may either lie in $W$ or $W'$ and similarly for $\tilde C\times (-1,0)$. Without loss of generality, we choose $\tilde C\times (0,1)\subset W'$. Now we can construct a loop homotopic to $\partial D_0$ with no cross with $p^{-1}(C)$. We may simply take $D_0\in N$ around $(0,0)$, whose boundary intersects $\tilde C$ at $t_1,t_2$. Then we choose two points $t'_1,t'_2\in \partial D_0\cap W'$ near $t_1$ and $t_2$ respectively. The small arc connecting $t'_1$ and $t'_2$ is taken to the small arc on $\partial D_0$, while the large arc connecting them is taken to be a curve nearly parallel to $\tilde C$. We denote this loops as $l_{t'_1,t'_2}$. It is homotopic to $(\partial D_0\cap \bar W')\cup [\tilde C\cap (\mathbb R^2-D_0)]$. The second part $[\tilde C\cap (\mathbb R^2-D_2)]$ is homotopic to $\partial D_0\cap \bar W$ since $\bar W-D_0$ is simply connected. By projecting back the loop $l_{t'_1,t'_2}$ to the torus, we obtain a loop that is homotopic to $\partial D_0$ on $T^2-{x_0}$ while its braid is not destroyed by the trajectory of EPs. This does not meet the annihilation condition. So a simple loop $C$ of EP trajectory must lie in the  nontrivial classes of $\pi_1(T^2,x_0)$, which are generated by meridians and longitudes.

We show that for trajectories $C$ with finite self-interactions, the image of $C$ should contain great circles of the torus. However, in this situation, a contractible loop like the one along the BZ boundary also meets the criterion of crossing all contractible loops in $\pi(T^2,x_0)$. If the loop contains finite number of self-crosses, then we parameterize the loop as a map $c: S^1\to T^2$ on the interval $[0,2\pi)$. The self intersecting points are ordered as $s_0<s_1<s_3\dots$. The total number of self. Then there is a maximal $i$ such that $c(s_i)=c(s_0)$. If the loop $c[s_0,s_i]$ contains any meridians or longitudes, we are done. If not, this loop is contractible and we can replace it by a constant $c'[s_0,s_i]=c(s_0)$. Choose $c'(s)=c(s)$ for $s\not\in [s_0,s_1]$, we obtain a new loop. Continue repeating this examination until all self-intersecting points are sorted, we obtain a simple contractible loop $C_1$. As before we lift this loop to the cover space $\mathbb R^2$. The lifted loop $\tilde C$ does not intersect with its translation counterparts, otherwise it must contain some meridians or longitudes. The lifted loop $C_1$ separates $\mathbb R^2$ into two connected components $W_1,W'_1$. Now examine $\tilde C\cap \bar W'_1$. This curve includes $\tilde C_1$ and those segments intersecting $\tilde C_1$ finite times, at a subset $\{s'_j\}\subset \{s_j\}$. As before, we start to look at the first intersecting point $s'_1$. If $\tilde C$ travels along $\tilde C_1$ after $s'_1$ or move towards the bounded component $W_1$, we do not need to do anything and look at the next intersecting point. If $\tilde C$ moves towards the unbounded component $W'_1$, it will intersect again with $\tilde C_1$ at some point $s'_j$. First consider $c(s'_j)\ne c(s'_1)$. Denote this curve as $w_{s'_1,s'_j}$. It may contain some self-intersections, which we can as before contract to a constant map. So $w_{s'_1,s'_j}$ can be treated as a simple curve. Together with the two arcs on $C_1$ connecting $s_1$ and $s'_j$, $w_{s'_1,s'_j}$ can form two simple loops. Each of them separates $\mathbb R^2$ into two connected components. We choose the one that intersects with the EP. We denote this simple loop as $\tilde  C_2$. In this way we eliminate all self-intersections of $\tilde C$ in the interior of $\tilde C_2$. If $c(s'_1)=c(s'_j)$, we may slight deform the curve around $s'_j$ so that it does not affect other intersections or connecting $\tilde C$ with its translations. Repeating doing this until we obtain some simple loop $\tilde C_j$ whose exterior does not intersect with $\tilde C$. Then we may as before construct a contractible loop $l$ on $\mathbb R^2-p^{-1}(x_0)$ that lies on the exterior of $\tilde C$ while does not intersect with its translation counterparts. Notice that $\tilde C_j$ is only piece-wise smooth. We might first construct segments of $l$ in the tubular neighborhood of each smooth piece of $\tilde C_j$. And then connect them together in the neighborhood of nonsmooth points. Projecting back $l$ to the torus we obtain the conclusion.

\subsection{Boundary Braid Illustration}
To better illustrate the braid nature of the boundary spectra, we include an alternative version of Fig.~\ref{fig:ham-d-3D-spectra}~(b), where we plot the spectrum as a three-dimensional plot, see Fig.~\ref{fig:3d-plot-boundary}.

\end{widetext}
~
\begin{figure}[h!]
    \centering
    \includegraphics[width=\linewidth]{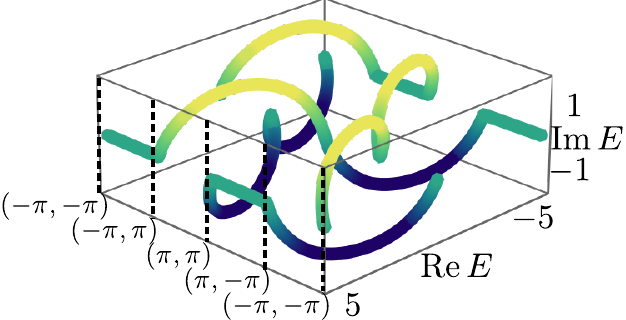}
    \caption{Three-dimensional plot of the boundary braid, alternative to Fig. 2 (b) in the main text. The color corresponds to the imaginary part of the eigenvalues.}
    \label{fig:3d-plot-boundary}
\end{figure}
\end{document}